\documentclass[prl,twocolumn]{revtex4}

\usepackage{graphicx}
\usepackage{amsmath}
\usepackage{amsfonts}
\usepackage{bm}

\begin{document}

%......Title and other stuff

\title{Electron-spin beat susceptibility of excitons in
semiconductor quantum wells}
\author{N. H. Kwong}
\author{S. Schumacher$^\dagger$}
\author{R. Binder}

\affiliation{College of Optical Sciences,
             University of Arizona,
             Tucson, Arizona 85721, USA}

\date{\today}

%\pacs{71.35.-y, 78.47.+p, 42.65.Sf}

\begin{abstract}
Recent time-resolved differential transmission and Faraday rotation
measurements of long-lived electron spin coherence in quantum wells
displayed intriguing parametric dependencies.  For their
understanding we formulate a microscopic theory of the optical
response of a gas of optically incoherent excitons whose constituent
electrons retain spin coherence, under a weak magnetic field applied
in the quantum well's plane. We define a spin beat susceptibility
and evaluate it in linear order of the exciton density. Our results
explain the many-body physics underlying the basic features observed
in the experimental measurements.

\end{abstract}

\maketitle

Spurred by prospects of applications in spintronics, the long-lived
electron spin coherence of excitations in semiconductor quantum
wells has been undergoing intensive investigation
\cite{wolf-etal.01,awschalom-etal.02,lau-etal.01}. Experimentally
one of the most direct and convenient ways to study this spin
coherence is through the measurement of its effects on the quantum
well's optical response. Nonlinear optics techniques such as
differential transmission (DT) and Faraday rotation (FR) of optical
probes have been used for this purpose
\cite{kikkawa-awschalom.98,samarth.04,palinginis-wang.04,shen-etal.05}.
Typically, the electron-hole excitation is produced in a pure spin
state aligned with the quantum well's growth axis in the presence of
a weak magnetic field applied along the well's plane (Voigt
geometry, Fig. 1a). Time-resolved DT and FR signals oscillating at
the electron spin Zeeman splitting frequency are then generated by
probe pulses at delay times spread over hundreds of picoseconds.

While the decay of the electron spin signals is by now well
understood (for a review, see e.g. \cite{zutic-etal.04}), other
fundamental parametric dependencies are still under active
investigation. Recent reports on the measurement (DT and FR) and
control of electron spin coherence in a pumped population of
excitons \cite{palinginis-wang.04,shen-etal.05,shen-etal.07} showed
intriguing dependencies on probe frequency and intensity, which
could reveal important information about the nonlinear optical
properties of the electron-spin coherent, but optically incoherent,
excitons. The experiments have been interpreted with
phenomenological models, but a microscopic theory would provide a
more in-depth understanding. The purpose of this letter is to
formulate a general microscopic theory of the nonlinear optical
susceptibility of a quantum well which carries a population of
electron-spin-coherent excitons.  Valid in linear order in both the
pumped exciton density and probe field amplitude, the theory
clarifies the physics of time-resolved DT and FR spin beats at the
low density limit.

\begin{figure}[b]
\includegraphics[scale=0.3,angle=0]{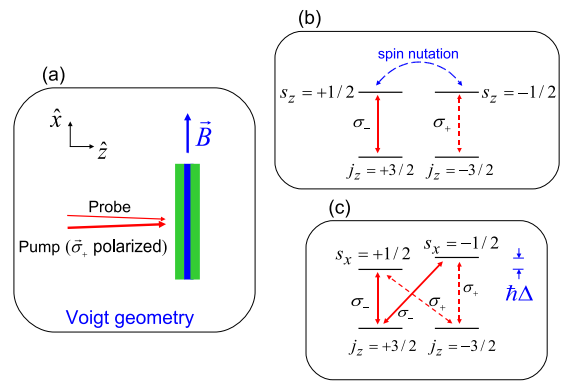}
\caption{\label{geometry} (a) Sketch of the Voigt geometry with
normal incidence light and a magnetic field ${\bf B}$ in the quantum
well plane. (b) Optical selection rules in the z-basis. The ${\bf
B}$-field induced coherence between the electron states in the
z-basis is indicated. (c) Optical selection rules using the z-basis
for holes and x-basis (eigenstates of magnetic field Hamiltonian)
for electrons. }
\end{figure}

Microscopic theories have been extensively developed for the
nonlinear response of quantum wells in the ultrafast ($\leq$ several
ps) regime. In conjunction with significant experimental efforts,
these theories have established exciton-exciton interactions as the
primary mechanism driving nonlinear optical effects (for recent
reviews, see e.g. \cite{chemla-shah.01,axt-kuhn.04,meier-etal.06}).
In particular, the microscopic processes underlying {\it
exciton-spin} beats in FR signals in this regime have been discussed
\cite{oestreich-etal.95a,sham.99}. But so far a microscopic theory
of the nonlinear optics of optically dephased,
electron-spin-coherent excitons on the 100 ps time scale is still
absent.

We consider specifically the experimental configuration sketched in
Fig. 1a. A circularly polarized, say $\sigma_+$, pump pulse at
normal incidence (along the $z$-axis) and spectrally close to the
lowest heavy hole exciton resonance creates an interband
polarization in a quantum well at low ambient temperature, e.g. 10K.
Within a time scale $\tau_R$ ($<$ 50 ps in GaAs at 10K) after the
pumping, the interband polarization partially re-radiates and
partially dephases and relaxes to yield a population of incoherent
heavy hole excitons with a distribution of center-of-mass momentum.
The hole spin inside a pumped exciton ($j_z = 3/2$ initially)
decoheres also within $\tau_R$, while the electron spin state ($s_z
= -1/2$ initially) stays pure for a long time, e.g. typically
hundreds of ps in GaAs quantum wells. A magnetic field applied along
an axis (the $x$-axis) in the plane of the quantum well splits the
two degenerate electron spin states quantized along the $x$-axis
(Fig. 1c) and, in the case at hand, drives a coherent oscillation of
spin population between the two electron spin states quantized along
the $z$-axis (Fig. 1b): with the electron created in $s_z = -1/2$,
the electron spin density matrix in the $z$-axis basis is
\begin{align}\label{spinmatrix-z}
{\hat \rho}^s_z (t) = \frac{1}{2}
\left(
\begin{array}{cc}
[1 + {\rm cos} \Delta (t - t_{\rm pu})] &  i {\rm sin} \Delta
(t - t_{\rm pu}) \\
- i {\rm sin} \Delta (t - t_{\rm pu}) & [1 - {\rm cos} \Delta (t
- t_{\rm pu})]
\end{array}
\right)
\end{align}
where $\hbar \Delta$ is the Zeeman splitting and $t_{\rm pu}$ is the
pump time.

Many properties of the exciton population can be described by the
one-exciton density matrix $\langle p^{s j \dag}_{\bf q} (t) p^{s'
j' }_{{\bf q}'} (t)
 \rangle $ where $p^{s j}_{\bf q}$ is the second
quantized annihilation operator for a 1s heavy hole exciton with
electron spin $s$ quantized along the $\hat z$-axis, hole spin $j$,
and center-of-mass in-plane momentum ${\bf q}$ (all momenta in this
paper lie in the quantum well's plane), and $\langle \cdot \rangle$
denotes an expectation value in a many-exciton state. For times
larger than $\tau_R$, when the state has decohered with respect to
hole spin and momentum, the one-exciton density matrix can to a good
approximation be written as an uncorrelated product of matrices in
the three state labels: $\langle p^{s j \dag}_{\bf q} (t) p^{s'
j'}_{{\bf q}'} (t) \rangle = [ {\hat \rho}^s_z ]_{s s'} \delta_{j
j'} \delta_{ {\bf q} {\bf q}'} f ({\bf q})$, where $f({\bf q})$ is
the slowly evolving momentum distribution . The relaxation dynamics
of $f({\bf q})$ has previously been investigated
\cite{piermarocchi-etal.96}. Based on their findings, we use a model
distribution, shown in Fig. 2b, which is thermal except for a
radiative loss correction at low momenta.

\begin{figure}
\includegraphics[scale=0.35,angle=-90]{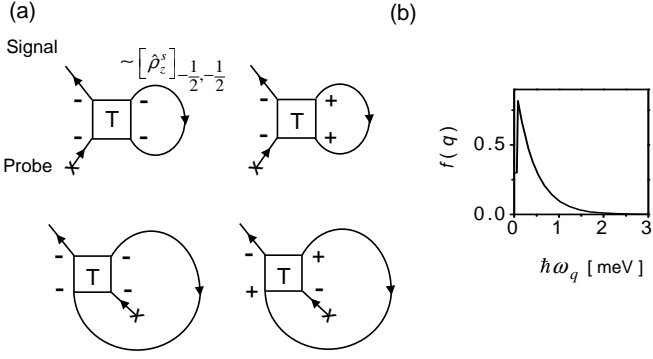}
\caption{\label{momentum-dist} (a) Diagrams representing the
excitonic scattering contributions to the third order interband
polarization. The internal directed line, labeled in one diagram by
$[ {\hat \rho}^s_z ]_{-1/2 -1/2}$, represents an optically
incoherent exciton. The box enclosing the letter 'T' denotes the
scattering matrix or T-matrix. The electron spin states ($+$ or $-$)
of the incoming and outgoing excitons of the scattering event are
marked. (b) A model momentum distribution of the relaxed excitons.}
\end{figure}

Microscopic many-particle theory comes into play when we try to
understand the measurement and manipulation of the electron spin
coherence of the exciton population.  Suppose a probe pulse
$E_{\sigma} (  t )$, circularly polarized ($\sigma = +$ or $-$) and
centered spectrally at $\omega$ and temporally at $t_{\rm pr}$, is
sent in to measure this coherence. It does so by inducing a 1s heavy
hole interband polarization which scatters off the existing spin
coherent exciton population. We use real-time Green's functions and
diagrammatic perturbation methods to derive equations of motion for
the probe-induced interband polarization in increasing orders of the
pump and probe intensities \cite{kwong-binder.00}. These equations
are derived with electrons and holes as degrees of freedom. They are
then expanded in an exciton basis and restricted to the heavy hole
1s subspace. The resulting equations -- our working equations -- are
driven by various exciton scattering processes and Pauli blocking.
(Pauli blocking turns out to be relatively unimportant here.) We
illustrate the lowest order scattering processes in Fig. 2a, showing
the various electron spin (in the $\hat z$-basis) combinations
explicitly. In each diagram, the line starting with a cross denotes
the 1s interband polarization created by the probe pulse, and the
open line a propagating exciton that recombines to give the signal
photon. The internal line represents a two-time correlation function
of the pumped exciton population $\langle p^{s j \dag}_{\bf q} (t)
p^{s' j' }_{{\bf q}'} (t')  \rangle = \langle p^{s j \dag}_{\bf q}
(t) p^{s' j' }_{{\bf q}'} (t)  \rangle {\rm exp} [i \omega_q ( t -
t' )]$, where the signs of the electron spin $s$ and $s'$ are marked
at the line's two ends, and $\hbar \omega_q$ is the energy of the
exciton with momentum ${\bf q}$. The scattering matrix
\cite{takayama-etal.02}, or T-matrix, is a ladder sum of repeated
exciton interactions to all orders and so can include the process of
biexciton formation. The excitonic T-matrix conserves the total
electron spin of the two scattering excitons, which implies, as
illustrated in Fig. 2a, that only the diagonal part of the electron
spin density matrix, Eq.~(\ref{spinmatrix-z}), is probed. In other
words, the optical probe measures the coherent oscillation of
exciton density between the two $\hat z$-quantized electron spin
states.

The susceptibility for the $\sigma$-(circularly)polarized
component of the probe is given by $\chi_{\sigma}  %$
%$\chi_{\sigma} (\omega , \tau )
\sim
%- \mu^\ast {\tilde \phi}
\int d t p_{\sigma} ( t ) E^\ast_{\sigma} ( t )$ where $p_{\sigma} (
t )$ is the probe-induced interband polarization. The susceptibility
depends parametrically on the pump-probe delay time $\tau = t_{\rm
pr} - t_{\rm pu}$ and the probe pulse's center frequency $\omega$:
$\chi_{\sigma} = \chi_{\sigma} (\omega , \tau )$. It gives the
time-resolved FR of an $x$-polarized probe and the DT of a
$\sigma$-polarized probe as \cite{sham.99} $\theta (\omega , \tau )
\sim {\rm Re} [ \chi^{}_{-} (\omega , \tau ) - \chi^{}_{+} (\omega ,
\tau ) ]$ and $\Delta T_{\sigma} (\omega , \tau ) \sim - {\rm Im} [
\chi^{}_{\sigma} (\omega , \tau ) ]$ respectively. We have
calculated the susceptibility to linear order in the pump-induced
density, which we call the third order susceptibility
$\chi^{(3)}_{\sigma} ( \omega , \tau )$.  Within our theory, it can
be written in the form
\begin{equation} \label{chi_total}
\chi^{ (3)}_{\pm} ( \omega, \tau ) = C ( \omega ) \pm D_{\rm spin}
(\omega) {\rm cos} \Delta \tau
\end{equation}
The response signals are similarly parameterized: $\theta (\omega ,
\tau ) = A_{\theta} (\omega) {\rm cos} \Delta \tau$, $\Delta T_{\pm}
(\omega , \tau ) = A (\omega) \pm A_T {\rm cos} \Delta \tau$, with
the amplitudes given by $A_{\theta} (\omega) \sim - {\rm Re} D_{\rm
spin} (\omega)$, $A_{T} (\omega) \sim - {\rm Im} D_{\rm spin}
(\omega)$, and $A (\omega) \sim - {\rm Re} C (\omega)$. The probe
pulses we use have durations of several ps, which is long compared
to the scattering duration (typically $<$ 1 ps in GaAs) but short
compared to the spin beat period ($\sim$ 70 ps in the cited
experiments). It is then instructive to approximately treat the
probe pulse as a continuous wave in the scattering calculation and
the oscillating electron spin population as frozen within the
duration of $p^{(3)}_{\sigma} (t)$. Solving the $p^{(3)}_{\sigma}$
equation in this approximation, we obtain the $\omega$-dependent
coefficients of $\chi^{(3)}_{\sigma}$ as
\begin{widetext}
\begin{eqnarray} \label{chi_non_beat}
C (\omega) \sim
 \sum_{\bf q}  \frac {f ( {\bf q} )} {(
L(\omega) )^2} & &  \left[ - T^{\pm \pm}_{{\bf q}/2, {\bf q}/2} (
\Omega_{\bf q} , {\bf q} ) - T^{\pm \pm}_{{\bf q}/2, -{\bf q}/2} (
\Omega_{\bf q} , {\bf q} ) - T^{\pm \mp}_{{\bf q}/2, {\bf q}/2} (
\Omega_{\bf q} , {\bf q} ) + \frac {1} {2} T^{\pm \mp}_{{\bf q}/2, -
{\bf q}/2} ( \Omega_{\bf q} , {\bf q} ) \right. \nonumber \\ & & +
\left.  ({\widetilde A}^{\rm PSF}_e ( {\bf q} ) + {\widetilde
A}^{\rm PSF}_h ( {\bf q} ) ) L(\omega)  \right]
\end{eqnarray}
\begin{equation} \label{chi_beat}
D_{\rm spin} (\omega) \sim
 \sum_{\bf q}  \frac {f ( {\bf q} )} { ( L(\omega) )^2} \left[ - T^{\pm
\pm}_{{\bf q}/2, {\bf q}/2} ( \Omega_{\bf q} , {\bf q} ) + T^{\pm
\mp}_{{\bf q}/2, {\bf q}/2} ( \Omega_{\bf q} , {\bf q} )- \frac {1}
{2} T^{\pm \mp}_{{\bf q}/2, - {\bf q}/2} ( \Omega_{\bf q} , {\bf q}
) + {\widetilde A}^{\rm PSF}_e ( {\bf q} ) L(\omega) \right]
\end{equation}
\end{widetext}
Here $T^{\sigma \sigma'}_{{\bf k}_f {\bf k}_i} (\Omega , {\bf Q} )$
denotes a two-exciton scattering (or T-matrix) element
\cite{takayama-etal.02} between an incoming state with relative
momentum ${\bf k}_i$ and an outgoing state with relative momentum
${\bf k}_f$, and $\hbar \Omega$ and ${\bf Q}$ are the conserved
total energy and momentum respectively. The superscripts $\sigma,
\sigma'$ label the polarization channel: $T^{++} (=T^{--})$ is the
T-matrix for co-polarized excitons, and $T^{+-} (=T^{-+})$ is that
for counter-polarized excitons. For each process both the incoming
and outgoing states contain a virtual (probe) exciton with zero
momentum and energy $\hbar \omega$ and a real (pump) exciton with
momentum ${\bf q}$ and energy $\hbar \omega_q = \varepsilon_x +
\hbar^2 q^2 / ( 2 M_x )$, $M_x$ being the exciton's mass. These give
a total energy of $\Omega_{\bf q} = \omega + \omega_q$ and a total
momentum of ${\bf q}$. $L(\omega)= \hbar \omega - \varepsilon_x + i
\gamma$ where $\gamma$ is the linewidth of the exciton resonance.
The phase space filling factors are given in terms of the momentum
space 1s exciton wavefunction $\phi ( {\bf k} )$ by ${\widetilde
A}^{\rm PSF}_{\alpha} ( {\bf q} ) = \sum_{\bf k} \phi ( {\bf k} ) |
\phi ( {\bf k}_{\alpha} ) |^2 / \sum_{\bf k} \phi ( {\bf k} )$ ,
where ${\bf k}_{\alpha} = {\bf k} - (m_{\alpha}/M_x) {\bf q}$, with
$\alpha = e , h$. The T-matrix elements are calculated by
diagonalization of the two-exciton Hamiltonian as explained in
\cite{takayama-etal.02}.

\begin{figure}
\includegraphics[scale=0.86]{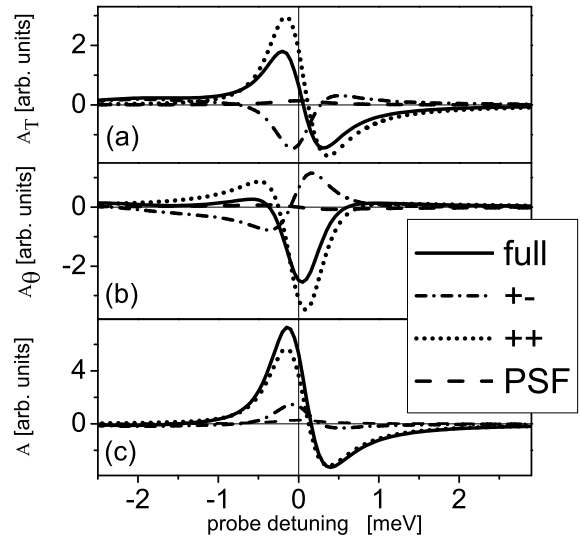}
\caption{\label{DT-FR} (a) and (b). The (signed) amplitudes of the
electron spin beat signals of a GaAs quantum well as a function of
probe frequency for a (+)-polarized pump: (a) differential
transmission (DT) of a (+)-polarized probe and (b) Faraday rotation
(FR) of a linearly polarized probe. (c) The non-beat part of the DT.
Also shown is the breakdown of each signal into its components:
phase space filling (dashed), exciton scatterings $T^{++}$ (dotted)
and $T^{+-}$ (dash-dotted).}
\end{figure}

\begin{figure}
\includegraphics[scale=0.96]{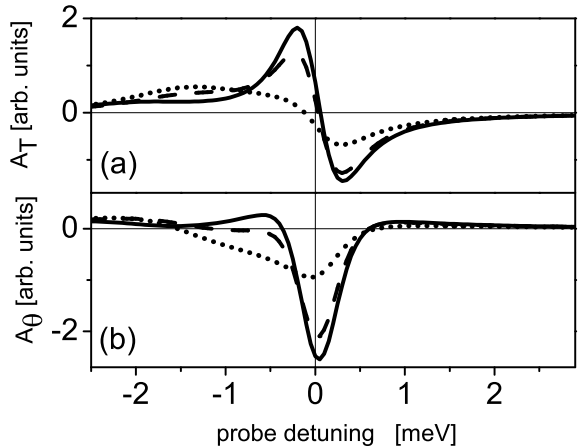}
\caption{\label{adjust} Sensitivity of the electron spin beat
signals to variations in the strength and/or resonance energy of the
bound biexciton. Solid: same as solid curves in Fig. 3, for which
the biexciton binding energy $\varepsilon_{bx}$ is calculated to be
1.8 meV. Dashed: $\varepsilon_{bx}$ is shifted to 1.3 meV, other
components being the same as for the solid curves. Dotted:
$\varepsilon_{bx}$ is shifted to 1.3 meV, and biexciton strength is
raised by a factor of 1.5.}
\end{figure}

Equation~(\ref{chi_total}) shows a simple dependency to pump-probe
polarization configurations: the co-polarized ($+$ pump and $+$
probe) and counter-polarized ($+-$) response signals share a common
$\tau$-independent part but have $180^\circ$-out-of-phase spin beat
parts. At the same time, Eqs. (\ref{chi_non_beat}) and
(\ref{chi_beat}) show that, even for the co-polarized configuration,
say, both $T^{++}$ and $T^{+-}$ contribute. In particular, biexciton
formation, present in $T^{+-}$, plays a part in the co-polarized
channel response. Moreover, Eq. (\ref{chi_total}) shows that the
non-beat part of the FR of a linearly-polarized probe vanishes.
These properties follow from our assumption of hole spin
thermalization, which produces all four exciton ($s_z, j_z$) spin
states in the relaxed exciton population even though the pump is
circularly polarized. These polarization dependencies are in obvious
contrast to those of the $\chi^{(3)}$ susceptibility in ultrafast
pump-probe spectroscopy (e.g. \cite{takayama-etal.04}), where
$T^{\sigma \sigma'}$ contributes only in the matching ($\sigma
\sigma'$) pump-probe polarization channel, yielding distinctly
different responses in the ($++$) and ($+-$) pump-probe
configurations \cite{sieh-etal.99}.

We have calculated $C(\omega)$ and $D_{\rm spin} (\omega)$ for the
exciton momentum distribution shown in Fig. 2b, with an exciton
linewidth of $\gamma = $0.35 meV and a biexciton dephasing of $2
\gamma$. In our calculation, the $p^{(3)}_{\sigma} (t)$ equation is
solved with 4-ps probe pulses, i.e., without assuming the probe to
be a continuous wave. The effect of the 'continuous-wave'
approximation is, however, small so that Eqs. (\ref{chi_non_beat})
and (\ref{chi_beat}) can be used to interpret the numerical results.
Fig. 3 shows (a) the spin beat amplitude $A_T (\omega)$ of the DT of
a (+)-polarized (co-polarized with the pump) probe, (b) the beat
amplitude $A_{\theta} (\omega)$ of the FR of a linearly polarized
probe, and (c) the non-beat part $A (\omega)$ of the DT. The
signals' spectral behaviors are basically products of those of the
T-matrix elements and one-exciton Lorentzians $[L(\omega)]^{-2}$.
The spectra of our finite-${\bf q}$ T-matrix here are qualitatively
similar to those of the zero-${\bf q}$ T-matrix for ultrafast
nonlinear optics discussed in \cite{takayama-etal.02}. A prominent
feature in Fig. 3 is the considerably larger magnitude of the
non-beat part of the DT signal compared to $A_T$ and $A_{\theta}$.
This can be understood from the breakdown into contributions from
various processes, also shown in the figure. For the spin beat
amplitudes, the contributions from $T^{++}$ (dotted lines) and
$T^{+-}$ (dash-dotted lines) largely counteract each other,
especially near the exciton resonance ($\hbar \omega \approx
\varepsilon_x$). This effect can be expected from the opposite signs
with which the two contributions appear in Eq. (\ref{chi_beat}) for
$D_{\rm spin} (\omega)$. In contrast, as seen from Eq.
(\ref{chi_non_beat}), $T^{++}$ and $T^{+-}$ tend to reinforce each
other in $C (\omega)$, which also contains an extra $T^{++}_{{\bf
q}/2, -{\bf q}/2}$.

The smallness of $A_T$ relative to the non-beating part of the DT
signal is in accord with experimental measurements of DT, especially
around zero probe detuning \cite{palinginis-wang.04,shen-etal.05}.
The relative phase (= $180^\circ$) between the beats in the DT in
the two pump-probe polarization configurations is also confirmed
\cite{shen-wang.07}. The $\omega$-dependence of $A_{\theta}$ shown
in Fig. 3 also agrees qualitatively with measurements
\cite{shen-etal.05}. We note that our theory is essentially
parameter-free (the input parameters being the electron and hole
masses, the background dielectric constant, and environmental
dephasings) with a model of a zero-width quantum well with two
bands. Bearing this model's limitations in mind, we think the
general agreement between our predictions and experiments is
sufficient to validate the physical points of our theory. More
accurate modeling of the experimental samples, such as including
effects of a finite well width and the light hole band, will be done
in the future. To get a rough sense of how these modeling advances
might change our results, we have repeated our calculations,
artificially varying the biexciton energy and strength from their
calculated values. These biexciton characteristics are known to
depend quite sensitively on the well width. From Fig. 4, one can see
the qualitative features of our theory are robust, but improved
sample modeling would be needed for precise predictions.

In summary, we have formulated a microscopic theory, based on
exciton interactions, for the electron spin beat components of
differential transmission and Faraday rotation signals of quantum
wells carrying a population of dephased, but electron spin coherent
excitons. This theory explains the basic features of recent
experimental results at the limit of low pump-induced density and
low probe intensity. The theory will be generalized to higher orders
in the probe intensity and to three-pulse configurations
\cite{shen-etal.07}.

We thank H. Wang, Y. Shen, and S. O'Leary for valuable discussions
and JSOP for financial support. \mbox{S. Schumacher} gratefully
acknowledges financial support by the DFG (SCHU~1980/3-1).

$^\dagger$ Present address: Physics Department, Heriot-Watt
University, Edinburgh EH14 4AS, UK

\end{document}